\documentclass[aps,pra,reprint,superscriptaddress]{revtex4-2}
\usepackage{amsmath}
\usepackage{amssymb}
\usepackage{amsfonts}
\usepackage{physics}
\usepackage{graphicx}
\usepackage{float}
\usepackage[utf8]{inputenc}

\begin{document}
\title{Pauli Weight Hamiltonian Term Selection for Optimized Machine Learning Based Quantum Error Mitigation}
\author{Fadhil Fatih Shiddiq}
\affiliation{Department of Physics, Faculty of Science and Mathematics, Bandung Institute of Technology, Bandung 40132, Jawa Barat, Indonesia }
\author{Darell Timothy Tarigan}
\affiliation{Department of Physics, Faculty of Science and Mathematics, Bandung Institute of Technology, Bandung 40132, Jawa Barat, Indonesia }
\author{Hadyan Luthfan Prihadi}
\affiliation{Research Center for Quantum Physics, National Research and Innovation Agency (BRIN), South Tangerang 15314, Indonesia.}
\author{Jusak S. Kosasih}
\affiliation{Theoretical High Energy Physics Group, Department of Physics, FMIPA, Institut Teknologi Bandung, Jl. Ganesha 10 Bandung, Indonesia.}
\author{Yanoar P. Sarwono}
\email{yano001@brin.go.id}
\affiliation{Research Center for Quantum Physics, National Research and Innovation Agency (BRIN), South Tangerang 15314, Indonesia.}
\author{Leong-Chuan Kwek}
\email{kwekleongchuan@nus.edu.sg}
\affiliation{Centre for Quantum Technologies, National University of Singapore, Singapore 117543, Singapore}
\affiliation{National Institute of Education, Nanyang Technological University, 1 Nanyang Walk, Singapore 637616}
\affiliation{MajuLab, CNRS-UNS-NUS-NTU International Joint Research Unit, UMI 3654, Singapore}
\affiliation{Quantum Science and Engineering Centre, Nanyang Technological University, Singapore}
\author{Freddy Permana Zen}
\email{fpzen@itb.ac.id}
\affiliation{Theoretical High Energy Physics Group, Department of Physics, FMIPA, Institut Teknologi Bandung, Jl. Ganesha 10 Bandung, Indonesia.}
\affiliation{Indonesia Center for Theoretical and Mathematical Physics (ICTMP), Institut Teknologi Bandung, Jl. Ganesha 10 Bandung, 40132, Indonesia. }

\begin{abstract}
    Machine learning provides a scalable solution for quantum error mitigation. However, the selection of appropriate Pauli strings for inclusion in training data remains a challenge. Current methods rely on heuristic or uniform random sampling, requiring data for every Pauli string in the Hamiltonian, a process that scales linearly with measurements and grows with system size. To address this, we introduce quantum error mitigation with prior knowledge of Pauli weights (Pauli weight quantum error mitigation (Pi-QEM)), a systematic framework that selects training observables based on Pauli weight. By leveraging the relationship between variance and locality in parameterized quantum circuits, Pi-QEM trains on a small subset of dominant, low-weight Pauli strings. In numerical simulations of molecular systems on a noisy IBM quantum backend, Pi-QEM reduces ground-state energy estimation error by up to 34.01\% using just a single dominant local observable, offering an efficient, scalable pathway for high-precision error mitigation on NISQ devices.
\end{abstract}

\maketitle
\section{Introduction}
Quantum computing \cite{Nielsen2010Quantum,Lloyd1996Universal,Feynman2018Simulating,Abrams1997Simulation,Georgescu2014Quantum} 
in the noisy intermediate-scale quantum (NISQ) era \cite{Bharti2022Noisy,McArdle2020Quantum,Cao2019Quantum,Bauer2020Quantum,Lau2022NISQ,Peruzzo2014variational,Farhi2014quantum,Hakim2026Quantum,Preskill2018Quantum} 
is fundamentally constrained by hardware imperfections, leading to noise-induced bias in the expectation values of computational algorithms \cite{Wang2021Noise,Ma2020Quantum,Georgopoulos2021Modeling,Perdomo2018Opportunities,Arute2019Quantum,Iryanti2025Connecting}. 
Although quantum error correction (QEC) \cite{Terhal2015Quantum,Chiaverini2004Realization,Knill2000Theory,Cory1998Experimental} 
promises fault-tolerant computation \cite{Shor1996Fault,Gottesman1998Theory}, its significant qubit overhead makes it impractical for near-term quantum devices \cite{Fowler2012Surface,Kivlichan2020Improved,Lee2021Even}. 
Consequently, quantum error mitigation (QEM) has emerged as an essential alternative to suppress computational errors via classical post-processing without demanding large additional qubit resources \cite{Cai2023Quantum,Kandala2019Error,Kim2023Scalable,Suzuki2022Quantum,Bultrini2023Unifying,Temme2017Error,Li2017Efficient}.

Conventional QEM techniques, such as zero-noise extrapolation (ZNE) \cite{Temme2017Error,Li2017Efficient,Giurgica2020Digital} 
and probabilistic error cancellation (PEC) \cite{Temme2017Error,Bravyi2021Mitigating}, have been successfully deployed to push the limits of near-term quantum processors. 
However, both conventional methods are fundamentally constrained by a sampling overhead that may scale exponentially with the overall circuit fault rate \cite{Takagi2023Universal,Tsubouchi2023Universal,Quek2024Exponentially}, 
severely hindering their practical scalability for deep circuits and large molecular Hamiltonians. 
To alleviate the prohibitive measurement costs and rigorous calibration requirements of traditional QEM, learning-based quantum error mitigation methods have emerged as a highly efficient alternative. 
In this approach, a classical machine learning model is trained on classically simulable data to learn the mapping between noisy and ideal expectation values, thereby removing the need for additional mitigation circuits during runtime \cite{Czarnik2025Improving,Strikis2021Learning,Czarnik2021Error,Kim2020Quantum,Liao2024Machine}. 
A detailed benchmarking study \cite{Liao2024Machine} assessed various statistical models and found that the random forest regression–based implementation of learning-based quantum error mitigation (RF-QEM)  consistently outperformed other method. its demonstrated superior mitigation accuracy compared to digital zero-noise extrapolation (ZNE) while also significantly reducing runtime overhead. 
However, a systematic selection strategy and criterion for identifying the most statistically informative observables remain elusive, let alone a method for determining an optimal subset of observables, meaning current approaches rely instead on unguided random sampling. 
In the context of variational quantum eigensolver (VQE) \cite{Peruzzo2014variational} for molecular Hamiltonians, this limitation significantly affects the efficiency of learning-based QEM methods, as different observables contribute unevenly to the statistical information contained in the training data. 
Incorporating uninformative observables leads to inefficient data generation and degrades the overall learning performance, while excluding highly informative ones can significantly degrade model performance. 
Crucially, a small and well-chosen subset of the total Pauli terms comprising the Hamiltonian can accurately mitigate the remaining unseen Pauli observables. 
A systematic selection criterion is therefore essential to ensure the scalability and effectiveness of RF-QEM in practical VQE applications.

In this work, we introduce Pauli weight Quantum Error Mitigation (Pi-QEM), a systematic observable selection framework that addresses this open problem. 
Pi-QEM identifies statistically informative Pauli observables based on the variance of their expectation values across the parameterized quantum circuits (PQCs) space, $\text{Var}_{\Theta}[f(\hat{P}_k)]$, which determines the expressivity of the training data used in RF-QEM. 
Importantly, this variance is fundamentally linked to the cost-concentration phenomenon and the emergence of barren plateaus in PQCs, where low-variance observables correspond to exponentially suppressed training signals \cite{Arrasmith2022Equivalence,Cerezo2021Cost}. 
Leveraging this connection, Pi-QEM approximates the Pauli weight-based selection through a Pauli-weight--based criterion, allowing the identification of informative observables without requiring explicit variance evaluation. 
By restricting the training dataset to the selected subset $\mathcal{P}_{\text{sub}} \subset \mathcal{P}$, the dataset complexity is reduced from $\mathcal{O}(MN)$ to $\mathcal{O}(M|\mathcal{P}_{\text{sub}}|)$, with $|\mathcal{P}_{\text{sub}}| \ll N$, while preserving mitigation accuracy. 
We apply Pi-QEM for the $\text{H}_2$ molecule, focusing on the role of observable selection. 
Our results show that a small subset of selected observables, often a single dominant term, is sufficient to capture the underlying noise structure of the full Hamiltonian. 
This leads to substantial improvements in mitigation performance, achieving up to a $34.01\%$ reduction in ground-state energy estimation error while significantly reducing the required training resources.

\section{Theory and Methods}
A Parameterized Quantum Circuit (PQC) can be expressed generally as a sequence of parameterized unitaries:
\begin{equation}
U(\theta) = \prod_{i=1}^{n(L+1)} U_i (\theta_i),
\end{equation}
where $U_i (\theta_i) = e^{-i\theta_i V_i} W_i$, $V_i$ is a Hermitian operator, $W_i$ is an unparameterized fixed unitary, $n$ denotes the number of qubits, and $L$ represents the number of layers in the PQC. The circuit is initialized in the computational basis state $|0\rangle^{\otimes n}$. Following the unitary evolution of the PQC, the expectation value of an observable $\hat{O} \equiv \otimes_{i=1}^n \hat{O}_i$ is estimated. 

Let $\mu = (\mu_1, \mu_2, \dots, \mu_n)$ denote the binary measurement outcome obtained by measuring each qubit in the computational basis. The measurement outcomes can be mapped to eigenvalues $\pm 1$ using the function:
\begin{equation}
f(\mu) = \prod_{i=1}^n (-1)^{\mu_i}.
\end{equation}
The expectation value of the observable is therefore obtained as the mean value of $f(\mu)$ over repeated measurements. In the ideal, error-free limit with infinite sampling, the expectation value of the observable can be written as:
\begin{equation}
f_{\text{ef}} (\hat{O}) = E[\hat{O}] = \text{Tr}[\rho(\theta) \hat{O}],
\end{equation}
where the quantum state produced by the PQC is:
\begin{equation}
\rho(\theta) = U(\theta) \left( |0\rangle\langle 0|^{\otimes n} \right) U^\dagger (\theta).
\end{equation}
The unitary operator $U$ implemented by the PQC can be decomposed as:
\begin{equation}
U(\theta) = U_L (\theta) U_{L-1} (\theta) \dots U_1 (\theta),
\end{equation}
where $U_i$ denotes the unitary operation corresponding to the $i$-th circuit layer. 

For example, the circuit of interest in this paper is an $n$-qubit PQC consisting of $L$ repetitive layers. These layers are parameterized by a set of rotation angles $\theta = \{\theta_1, \theta_2, \dots, \theta_{n(L+1)}\}$, where each parameter corresponds to a single-qubit rotation gate $R_x$, forming the set of rotation operations:
\begin{equation}
R(\theta) = \{R_1 (\theta_1 ), R_2 (\theta_2 ), \dots, R_{n(L+1)} (\theta_{n(L+1)} )\}.
\end{equation}
As illustrated in Fig. 1, each circuit layer $U_L (\theta)$ consists of a set of single-qubit $R_Y$ rotations followed by a linearly connected chain of CNOT gates acting between neighboring qubits. This circuit architecture corresponds to a hardware-efficient ansatz (HEA), which is widely used in the Variational Quantum Eigensolver (VQE) algorithm to approximate the ground-state energy of molecular systems \cite{Darmawan2025Improving, Ramadhan2026HardwareEfficient, Ramadhan2026Expressibility}.

\begin{figure}[b]
    \centering
    \includegraphics[width=\columnwidth]{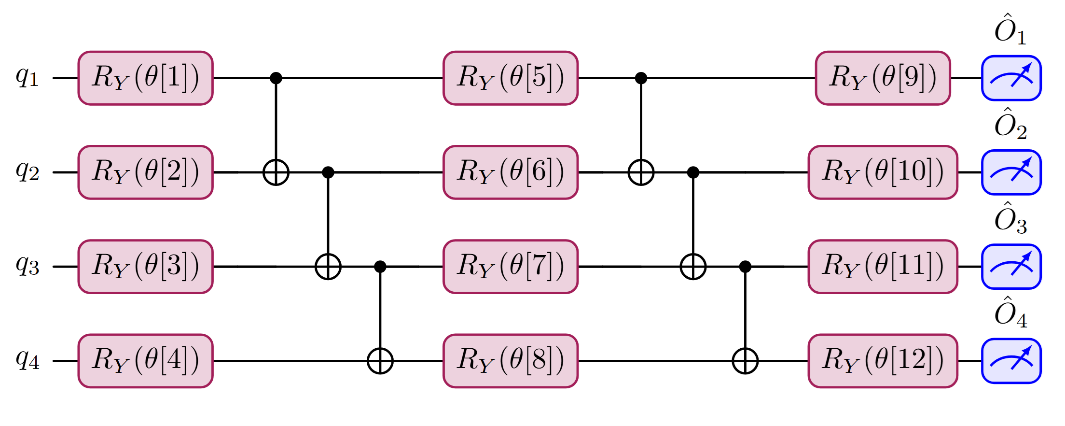}
    \caption{Schematic of a four-qubit, two-layer parameterized quantum circuit (PQC) based on a hardware-efficient ansatz (HEA). Each layer consists of single-qubit rotation gates $R_Y (\theta)$, followed by a linear chain of CNOT gates that entangle neighboring qubits. The circuit evolution concludes with measurements of the Pauli observables $\hat{O}_1$--$\hat{O}_4$}
    \label{fig:pqc_hea}
    
\end{figure}
\begin{figure*}[t]
    \centering
    \includegraphics[width=1.5\columnwidth]{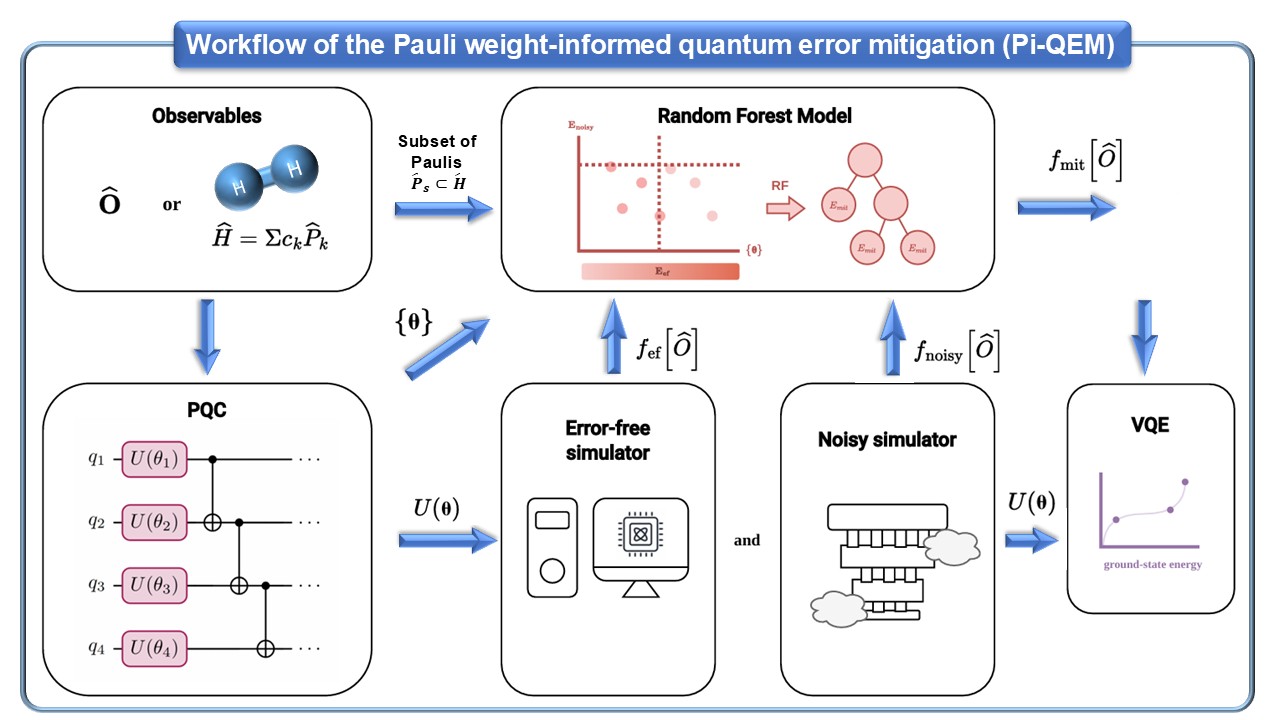}
    \caption{End-to-end workflow of the Pauli weight quantum error mitigation (Pi-QEM) framework. A selected subset of Pauli observables $\hat{P}_s \subset \hat{H}$ is combined with sampled PQC parameters $\{\theta\}$. The circuit is evaluated on both ideal ($f_{\text{ef}}$) and noisy ($f_{\text{noisy}}$) simulators to generate training pairs. The trained model maps noisy inputs to mitigated expectation values $f_{\text{mit}} [\hat{O}]$, enabling accurate ground-state energy reconstruction in the VQE.}
    \label{fig:pi_qem_workflow}
\end{figure*}

When the PQC is subjected to noise, the ideal unitary evolution is mapped by a noise channel $\mathcal{E}$ into a sequence of completely positive trace-preserving (CPTP) maps. In this framework, the entire noisy circuit can be represented as a composite quantum operation:
\begin{equation}
\mathcal{U} = \mathcal{U}_L \circ \mathcal{U}_{L-1} \circ \dots \circ \mathcal{U}_1,
\end{equation}
where $\mathcal{U}_i$ denotes the noisy quantum operation associated with the $i$-th circuit layer. In practice, each noisy layer is typically modelled as an ideal unitary evolution followed immediately by a noise channel. Thus, the operation of the $l$-th layer can be written as:
\begin{equation}
\mathcal{U}_l = \mathcal{E}_l \circ [\mathcal{U}_l ],
\end{equation}
where the super-operator
\begin{equation}
[\mathcal{U}_l](\cdot) \equiv U_l (\cdot) U_l^\dagger
\end{equation}
represents the ideal unitary transformation acting on the density matrix, and $\mathcal{E}_l$ denotes the noise channel associated with that layer. 

The noise channel $\mathcal{E}_l$ is a CPTP map that captures physical errors occurring in the quantum device, such as decoherence, relaxation, or gate imperfections. Starting from the initial state $|0\rangle^{\otimes n}$, the final noisy state of the quantum circuit can be expressed as:
\begin{equation}
\rho_{\mathcal{E}} = \mathcal{U} \left( |0\rangle \langle 0|^{\otimes n} \right).
\end{equation}

\begin{figure}[b]
    \centering
    \includegraphics[width=\columnwidth]{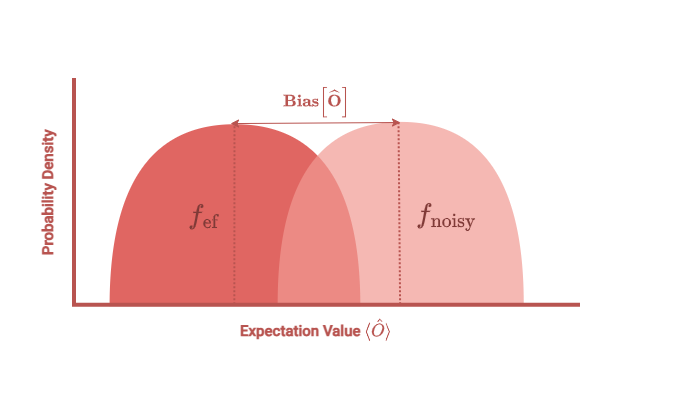}
    \caption{Illustration of noise-induced bias in quantum measurements. The probability distributions of the ideal expectation value ($f_{\text{ef}}$) and the noisy outcome ($f_{\text{noisy}}$) are shown. The shift between their peaks defines the systematic bias $\text{Bias}[\hat{O}]$, arising from the hardware noise channels.}
    \label{fig:noise_bias}
\end{figure}

Expanding the composite map explicitly gives:
\begin{equation}
\rho_{\mathcal{E}} = \mathcal{E}_L \left( U_L \dots \mathcal{E}_1 \left( U_1 |0\rangle \langle 0|^{\otimes n} U_1^\dagger \right) \dots U_L^\dagger \right).
\end{equation}
The expectation value of an observable $\hat{O}$ measured from this noisy state is then given by:
\begin{equation}
f_{\mathcal{E}} (\hat{O} ) = E_{\mathcal{E}} [\hat{O}] = \text{tr} [\rho_{\mathcal{E}} (\theta) \hat{O} ].
\end{equation}
This expression represents the experimentally accessible measurement outcome obtained from the noisy quantum circuit, which generally deviates from the ideal expectation value due to the accumulated effect of noise throughout the circuit. Consequently, the expectation value is biased, as illustrated in Fig. 2, where the bias is defined as:
\begin{equation}
\text{Bias}[\hat{O}] = |f_{\mathcal{E}} (\hat{O}) - f_{\text{ef}} (\hat{O})|.
\end{equation}

To mitigate the bias introduced by noise in quantum circuits and to learn the mapping between noisy and error-free expectation values, we construct a regression-based Pauli weight quantum error mitigation (Pi-QEM) framework illustrated in Fig. 3.

In noisy quantum devices, the expectation value of an observable $\hat{O}$ measured from a circuit subjected to a noise channel $\mathcal{E}$ deviates from its ideal value. Denoting the ideal expectation value as $f(\hat{O})$ and the noisy measurement outcome as $f_{\mathcal{E}} (\hat{O})$, the deviation can be interpreted as a noise-induced bias. In general, the mapping between the noisy and ideal expectation values depends on the PQC configuration and exhibits a non-linear relationship. Consequently, simple linear correction schemes are insufficient to capture this dependence. 

Instead, a supervised learning model that approximates the non-linear mapping between noisy measurements and their corresponding error-free values is constructed. The regression model is implemented using random forest regression, an ensemble learning method that aggregates the predictions of multiple regression trees to improve generalization and reduce overfitting \cite{Burkov2019hundred, James2013introduction}. Let $\{T_k \}_{k=1}^K$ denote an ensemble of $K$ regression trees. For a given input feature vector $x_i$, each tree produces an independent prediction $T_k (x_i)$. The mitigated expectation value is then obtained by averaging the predictions of all trees:
\begin{equation}
f_{\text{mit},i} (\hat{O} ) = \frac{1}{K} \sum_{k=1}^K T_k (x_i ).
\end{equation}

This ensemble averaging reduces the variance inherent in individual regression trees, allowing the model to approximate non-linear relationships between the noisy circuit outputs and their ideal expectation values. 

Each regression tree is constructed through the recursive partitioning of the feature space. At each node, the algorithm selects a feature $j$ and a split threshold $s$ that minimize the residual sum of squares (RSS):
\begin{equation}
\text{RSS} = \sum_{i:x_{ij} \le s} \left( y_i - \bar{y}_{R_{\text{left}}} \right)^2 + \sum_{i:x_{ij} > s} \left( y_i - \bar{y}_{R_{\text{right}}} \right)^2,
\end{equation}
where $y_i = f_{\text{ef},i} (\hat{O})$ denotes the target value corresponding to the ideal expectation value, while $\bar{y}_{R_{\text{left}}}$ and $\bar{y}_{R_{\text{right}}}$ represent the mean target values within the left and right partitions generated by the split, respectively.

By recursively applying this splitting procedure, each tree constructs a piecewise constant approximation of the underlying non-linear mapping between the noisy measurement outcomes and their corresponding ideal expectation values. 

\begin{figure}[b]
    \centering
    \includegraphics[width=\columnwidth]{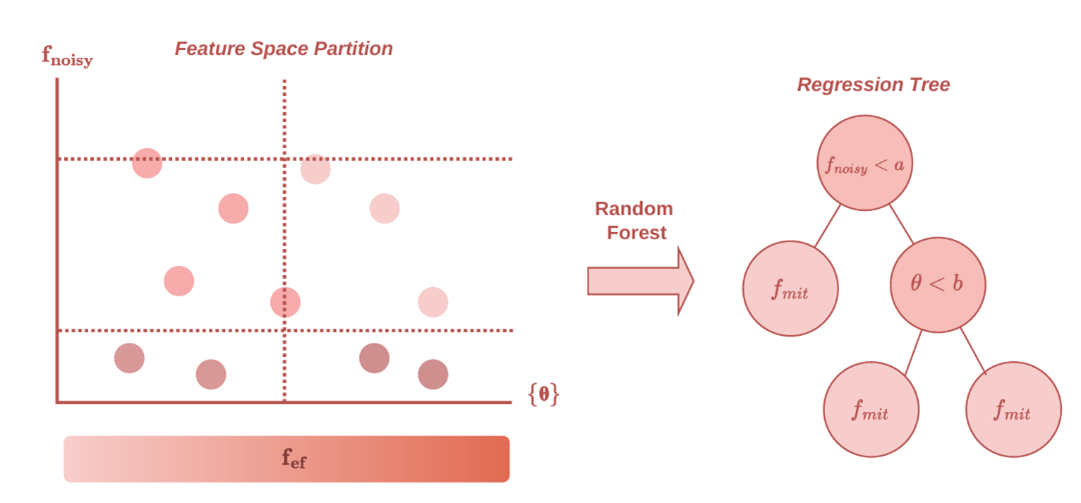}
    \caption{Conceptual illustration of the random forest regression process in the machine learning-based QEM method. The left panel shows recursive partitioning of the feature space, comprising noisy measurements ($f_{\text{noisy}}$) and circuit parameters ($\{\theta\}$), to minimize the residual sum of squares (RSS) with respect to the ideal target ($f_{\text{ef}}$). The right panel depicts the resulting decision tree mapping to mitigated outputs ($f_{\text{mit}}$).}
    \label{fig:rf_regression_qem}
\end{figure}

The training dataset is defined as $\mathcal{D} = \{(x_i, f_{\text{ef},i}(\hat{O}))\}_{i=1}^M$, where $M$ denotes the number of samples, $x_i$ is the input feature vector, and $f_{\text{ef},i}(\hat{O})$ is the corresponding error-free expectation value obtained from an ideal simulation. The feature vector is given by $x_i = [f_{\mathcal{E},i}(\hat{O}), \theta_{i,1}, \theta_{i,2}, \dots, \theta_{i,n(L+1)}]$, where $f_{\mathcal{E},i}(\hat{O})$ denotes the noisy expectation value under the noise channel $\mathcal{E}$, and $\theta_i$ represents the PQC rotation parameters. The inclusion of these circuit parameters enables the model to capture the dependence of the noise-induced bias on the parameter space. 

The dataset is generated through the following four-step procedure. First, a set of parameter vectors $\{\theta_i\}_{i=1}^M$ is sampled to explore the PQC configuration space. Second, for each sampled configuration, the circuit is evaluated in an ideal simulator to obtain the ground-truth expectation value $f_{\text{ef},i}(\hat{O})$. Third, the same circuit is executed under the noise channel $\mathcal{E}$ to produce the corresponding noisy expectation value $f_{\mathcal{E},i}(\hat{O})$. Finally, the noisy expectation value is combined with the associated PQC parameters to form the feature vector $x_i$, while the ideal expectation value is assigned as the target label. The regression model is then trained to learn the mapping $f_{\mathcal{E}}(\hat{O}) \to f_{\text{ef}}(\hat{O})$, allowing for the direct prediction of mitigated expectation values from noisy measurements. 

The performance of the model is quantified using the mean absolute error (MAE), defined as:
\begin{equation}
\text{MAE}(\hat{O}) = \frac{1}{M} \sum_{i \in \mathcal{D}} |f_{\text{mit},i}(\hat{O}) - f_{\text{ef},i}(\hat{O})|,
\end{equation}
which measures the average deviation between the mitigated predictions and the corresponding ideal expectation values. Minimizing this metric ensures that the predicted expectation values remain as close as possible to the true error-free results. 

During training, the Pi-QEM learns the non-linear relationship between noisy measurement outcomes and their corresponding ideal values through an ensemble of regression trees. Each tree independently approximates the mapping between the input features and the target expectation values, and the ensemble prediction is obtained by averaging the outputs across all trees. This aggregation reduces the variance associated with individual regression trees while preserving the model's ability to capture non-linear dependencies in the data.

Up to this point, the mitigation procedure has been formulated for the expectation value of a single Pauli observable $\hat{O}$. However, in practical applications such as VQE, the quantity of interest is the expectation value of a molecular Hamiltonian rather than a single observable. The Hamiltonian is typically expressed as a linear combination of Pauli strings obtained from fermion-to-qubit transformations, such as the Jordan–Wigner (JW), Bravyi–Kitaev (BK), and Parity (P) mappings. The qubit Hamiltonian is written as:
\begin{equation}
\hat{H} = \sum_{k=1}^N c_k \hat{P}_k,
\end{equation}
where $c_k \in \mathbb{R}$ are scalar coefficients determined by the molecular integrals, and $\hat{P}_k$ denotes Pauli string operators acting on the qubit register. The expectation value of the Hamiltonian with respect to a parametrized quantum state $|\text{\ensuremath{\psi}}(\theta)\rangle$ is then given by:
\begin{equation}
E(\theta) = \sum_{k=1}^N c_k f(\hat{P}_k).
\end{equation}
Consequently, accurate energy estimation requires reliable evaluation of the expectation values of all Pauli observables $\{\hat{P}_k\}_{k=1}^N$. 

Applying the mitigation procedure independently to each observable would require training $N$ separate regression models, which becomes computationally inefficient as the Hamiltonian size increases. Instead, we construct a generalized Pi-QEM model capable of mitigating all observables simultaneously. To achieve this, we extend the training dataset to include samples associated with the entire set of Pauli observables. Let
\begin{equation}
\mathcal{P} = \{\hat{P}_1, \hat{P}_2, \dots, \hat{P}_N\}
\end{equation}
denote the set of Pauli strings appearing in the Hamiltonian. The global training dataset is then defined as:
\begin{equation}
\mathcal{D}_{\text{global}} = \bigcup_{k=1}^N \{(x_{i,k}, f_{\text{ef},i}(\hat{P}_k))\}_{i=1}^M,
\end{equation}
where $f_{\text{ef},i}(\hat{P}_k)$ represents the ideal expectation value obtained from noiseless simulation, and $x_{i,k}$ denotes the corresponding feature vector associated with the $i$-th circuit configuration and the $k$-th observable. To enable the regression model to distinguish between different observables within the unified dataset, the feature vector is augmented with a representation of the Pauli string operator. The input features are therefore defined as:
\begin{equation}
x_{i,k} = [f_{\mathcal{E},i}(\hat{P}_k), \theta_{i,1}, \theta_{i,2}, \dots, \theta_{i,n(L+1)}],
\end{equation}
where $f_{\mathcal{E},i}(\hat{P}_k)$ is the noisy expectation value obtained from the circuit subjected to the noise channel $\mathcal{E}$, and $\theta$ denotes the set of PQC parameters defining the circuit state. This representation allows the regression model to incorporate observable-dependent noise behavior within a single unified framework. 

Training the RF model on $\mathcal{D}_{\text{global}}$ enables the ensemble of regression trees to learn the mapping between noisy and ideal expectation values across a larger and more expressive feature space. In particular, the model can capture correlations between circuit parameters, measurement noise, and the structure of the Pauli observables themselves. This shared representation allows the mitigation model to exploit common noise patterns across different observables measured within the same quantum circuit. 

Once the model has been trained, it can be applied to noisy expectation values obtained during the VQE optimization procedure. For each Pauli observable $\hat{P}_k$, the model predicts a mitigated expectation value $f_{\text{mit}}(\hat{P}_k)$. The mitigated estimate of the ground-state energy is then computed as:
\begin{equation}
E_{\text{mit}}(\theta) = \sum_{k=1}^N c_k f_{\text{mit}}(\hat{P}_k).
\end{equation}
This global mitigation strategy eliminates the need to train a separate regression model for every observable in the Hamiltonian while simultaneously improving robustness by leveraging shared noise characteristics across the entire quantum circuit.

We now select a Pauli weight-based observable for identifying subset $\mathcal{P}_{\text{sub}} \subset \mathcal{P}$ because Pi-QEM evaluates the statistical expressivity of each Pauli observable across the PQC parameter space and selects the most informative subset for Pi-QEM model training. In practice, constructing an effective Pi-QEM model means that every Pauli observable present in the Hamiltonian is not necessarily evaluated. Instead, a reduced training dataset can be constructed by selecting a subset of Pauli strings that retains the essential statistical structure of the full dataset. 

Let $\mathcal{P} = \{\hat{P}_1, \hat{P}_2, \dots, \hat{P}_N\}$ denote the complete set of Pauli observables appearing in the Hamiltonian. We define a reduced subset
\begin{equation}
\mathcal{P}_{\text{sub}} \subset \mathcal{P}
\end{equation}
containing only the selected Pauli strings. The corresponding reduced dataset is then given by:
\begin{equation}
\mathcal{D}_{\text{sub}} = \bigcup_{\hat{P}_k \in \mathcal{P}_{\text{sub}}} \{(x_{i,k}, f_i(\hat{P}_k))\}_{i=1}^M,
\end{equation}
which reduces the dataset size from $M \cdot N$ to $M \cdot |\mathcal{P}_{\text{sub}}|$. The objective of the subset selection procedure is to determine the subset $\mathcal{P}_{\text{sub}}$ that yields a mitigation model whose predictions closely approximate those obtained using the full observable set $\mathcal{P}$. 

Let $f_{\text{mit}}^{\text{sub}}(\hat{P}_k)$ denote the prediction produced by the model trained on the reduced dataset $\mathcal{D}_{\text{sub}}$. An optimal subset is defined as one that minimizes the MAE between the mitigated expectation value of $\hat{P}_k$ and its corresponding error-free expectation value:
\begin{equation}
\text{MAE}_{\text{sub}} = \frac{1}{M} \sum_{\theta \in \Theta} |f_{\text{mit}}^{\text{sub}}(\hat{P}_k) - f_{\text{ef}}(\hat{P}_k)|,
\end{equation}
where $\Theta = \{\theta_i\}_{i=1}^M$ denotes the set of sampled circuit parameter vectors. Minimizing this quantity ensures that the reduced model faithfully reproduces the predictions of the conventional model while requiring significantly fewer Pauli observables during dataset generation.

To identify the most informative Pauli observables for constructing the reduced subset $\mathcal{P}_{\text{sub}}$, we introduce a Pauli weight-based metric. First, consider the variance obtained from the expectation values of a Pauli observable evaluated at different circuit parameter configurations. Given a set of sampled parameters $\Theta$, the variance of these expectation values $f(\hat{P}_k)$ across the parameter samples is defined as:
\begin{equation}
\text{Var}_{\Theta} [f(\hat{P}_k)] = \frac{1}{M} \sum_{\theta \in \Theta} \left( f(\hat{P}_k) - \overline{f(\hat{P}_k)} \right)^2,
\end{equation}
where
\begin{equation}
\overline{f(\hat{P}_k)} = \frac{1}{M} \sum_{\theta \in \Theta} f(\hat{P}_k)
\end{equation}
denotes the mean expectation value across the sampled parameter set. This quantity characterizes how strongly the observable varies across the PQC parameter space and therefore quantifies the variability present in the training dataset. In the limit $M \to \infty$, the variance can be expressed in terms of expectations over the parameter distribution as:
\begin{equation}
\text{Var}_{\Theta} [f(\hat{P}_k)] = E_{\Theta} [f(\hat{P}_k)^2].
\end{equation}
$E_{\Theta} [f(\hat{P}_k)^2]$ captures the average magnitude of the signal generated by the observable $\hat{P}_k$ over the region of parameter space explored by the PQC.

The physical meaning of $\text{Var}_{\Theta} [f(\hat{P}_k)]$ as a selection metric is grounded in the cost-concentration phenomenon of PQC. The variance of the cost function difference between two independently sampled parameter points can be bounded as:
\begin{equation}
\text{Var}_{\Theta} [f(\hat{P}_k)] \le m^2 L_{\text{max}} F(n),
\end{equation}
where $m$ is the dimension of the parameter space, $L_{\text{max}}$ denotes the maximum length scale of the parameter space arising from the periodicity of the variational parameters, and $F(n)$ characterizes the concentration behavior of the landscape. This bound implies that $F(n)$ directly governs the scaling of $\text{Var}_{\Theta} [f(\hat{P}_k)]$, thus establishing a connection between observable variance and barren plateau formation \cite{Arrasmith2022Equivalence}. 

For global Pauli observables, $F(n)$ scales unfavorably with system size, leading to an exponential concentration of $f(\hat{P}_k)$ around its mean and consequently a vanishing $\text{Var}_{\Theta} [f(\hat{P}_k)]$. In contrast, for local observables, $F(n)$ exhibits at most polynomial scaling with system size, preventing concentration and preserving a non-trivial variance across the parameter space \cite{Cerezo2021Cost}.

\begin{figure}[b]
    \centering
    \includegraphics[width=\columnwidth]{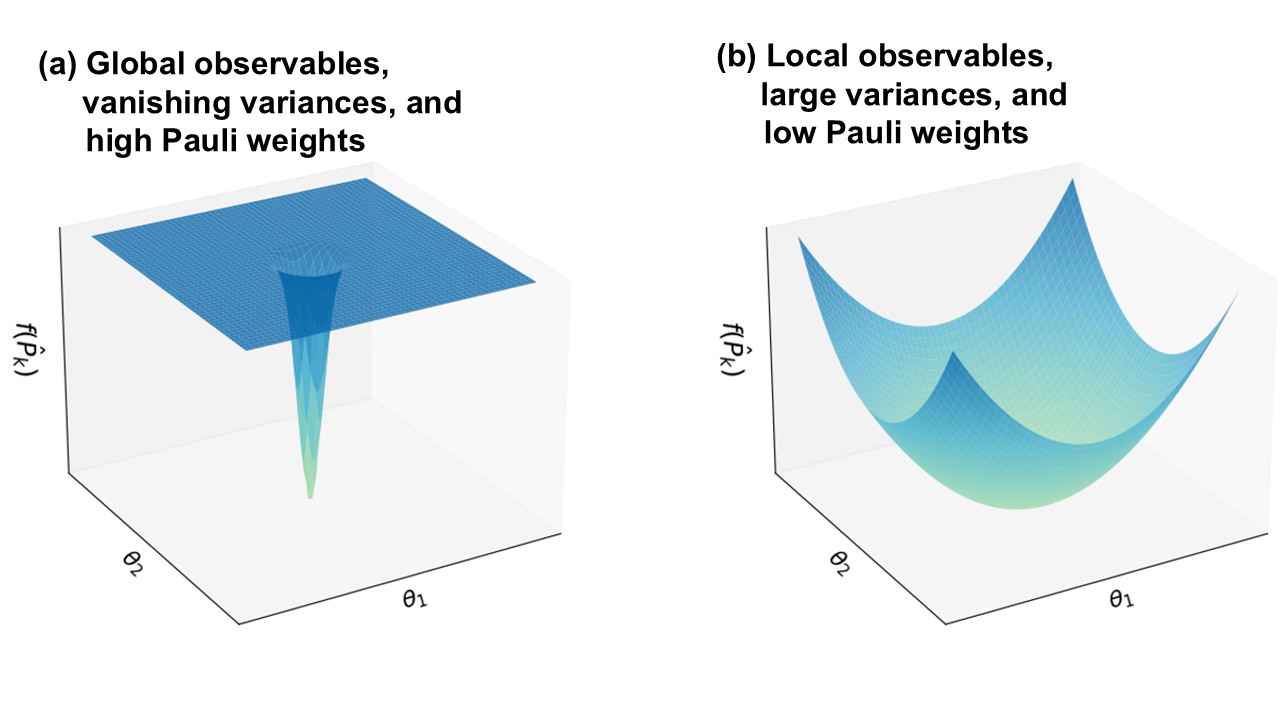}
    \caption{Landscapes of expectation values for different observable types, where $\theta_1$ and $\theta_2$ represent two variational parameters. (a) For global observables, the landscape exhibits a barren plateau, where the gradient vanishes along nearly all directions and the expectation values concentrate around a narrow range. (b) For local observables, the landscape remains free of barren plateaus; consequently, the cost function does not exhibit concentration, resulting in a comparatively larger variance of the expectation values.}
    \label{fig:landscapes_barren_plateau}
\end{figure}
As defined in Eq. (29), the variance of each Pauli observable quantifies how strongly its expectation value varies across the PQC parameter space, and therefore serves as a measure of its statistical contribution to the training dataset. Observables with larger variances generate more informative training signals, while those with vanishing variances contribute negligibly to the learning process. However, directly evaluating this variance for every Pauli term requires repeated circuit executions over numerous parameter configurations, which becomes computationally expensive as the system size grows. 

\begin{figure*}[t]
    \centering
    \includegraphics[width=2.0\columnwidth]{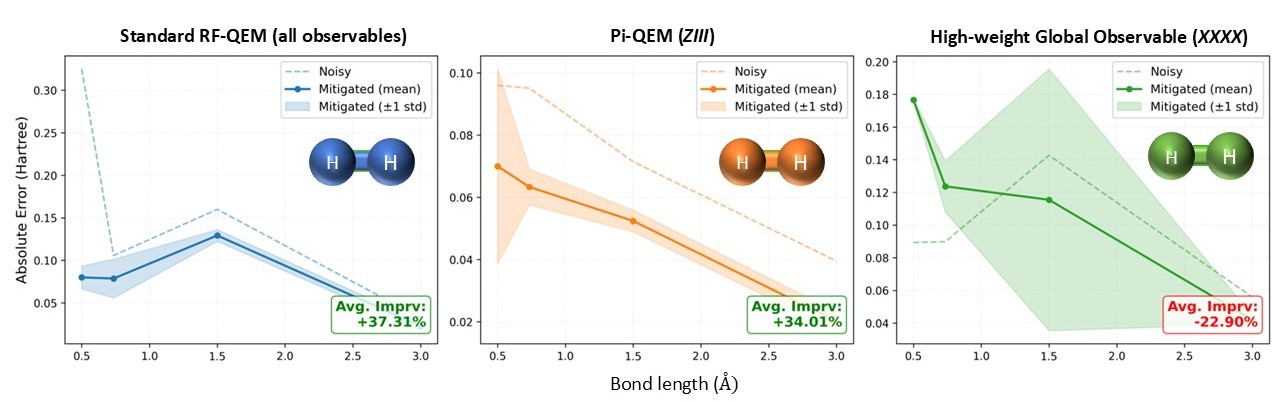}
    \caption{Absolute ground-state energy error (Hartree) vs. $\text{H}_2$ bond length (Å). The panels compare noisy (dashed) and mitigated (solid) results obtained from models trained on three distinct subsets: all Pauli observables (Standard RF-QEM), a low-weight local observable ($ZIII$, Pi-QEM), and a high-weight global observable ($XXXX$) that fails the criterion. The Pi-QEM model for the single low-weight local observable $ZIII$ (34.01\%) performs comparably to the standard RF-QEM model trained on all fifteen Pauli strings of the full Hamiltonian (37.13\%). Average improvement annotations (Avg. Imprv) are shown. Applying mitigation to unselected observables degrades performance.}
    \label{fig:h2_energy_error}
\end{figure*}

Importantly, barren plateau theory provides a structural explanation for this behavior. It predicts that the variance is strongly determined by the locality of the observable. Specifically, global Pauli strings exhibit exponentially suppressed variance, whereas local observables retain significantly larger variance across the parameter space. This implies that the ranking induced by Eq. (29) is not arbitrary, but is largely governed by the Pauli weight of each term. Consequently, this allows us to replace an explicit, data-driven selection criterion with a structure-based alternative. 

In particular, we approximate the variance-based selection using the Pauli weight $w(P_k)$, defined as the number of non-identity single-qubit operators in a Pauli string. Since the Pauli weight directly reflects locality, it provides a simple and scalable way to identify observables with high statistical relevance without explicitly computing their variance. We define Pi-QEM as an observable selection framework that identifies statistically informative Pauli terms based on their variance across the PQC parameter space, and implements this criterion in practice through a Pauli-weight–based approximation. The subset of dominant observables is therefore defined as:
\begin{equation}
\mathcal{P}_{\text{sub}} = \{P_k \in \mathcal{P} \mid w(P_k) \le w_{\text{max}}\},
\end{equation}
where $w_{\text{max}}$ controls the locality of the retained observables. 

When multiple observables share the same Pauli weight, a fixed number $s$ is selected uniformly at random within each weight group, allowing for flexibility in the subset size. In this way, Pi-QEM preserves its core principle of selecting statistically informative observables, while reducing the selection procedure to a simple operation on the Hamiltonian structure.

\section{Results and Discussion}
As a test system, we construct three models of the JW transformed $\text{H}_2$ molecule Hamiltonian, each trained on a distinct subset of data. The first is the standard RF-QEM model trained on all 15 Pauli strings of the full Hamiltonian. The second is the Pi-QEM model, where the training subset $\mathcal{P}_{\text{sub}}$ is defined using a Pauli-weight–based selection rule that targets a single dominant observable, $ZIII$, which is a low-weight (local) observable. The third model is trained on a high-weight global observable $XXXX$, representing a deliberately uninformative baseline. 

The training dataset consists of pairs of ideal expectation values from noise-free simulations and noisy expectation values obtained from simulations on the \texttt{ibm\_fake\_athens} backend \cite{Javadi2024Quantum}, which emulates realistic noise profiles of superconducting quantum hardware. 

Fig. 6 presents the ground-state energy estimation for the $\text{H}_2$ Hamiltonian across a range of bond lengths. The reported average improvement corresponds to the mean reduction in absolute energy error between the noisy and mitigated results over all bond lengths. The Pi-QEM result, corresponding to the selection of the single Pauli-weight local observable $ZIII$, achieves an average error reduction of 34.01\% relative to the noisy baseline. This performance is comparable to that of the standard RF-QEM model trained on all fifteen Pauli strings of the full Hamiltonian (37.13\%). In contrast, training on a high-weight global observable such as $XXXX$ leads to a 22.90\% performance degradation, demonstrating that observables failing the Pi-QEM selection criterion provide little informative structure for learning.

\begin{figure}[b]
    \centering
    \includegraphics[width=\columnwidth]{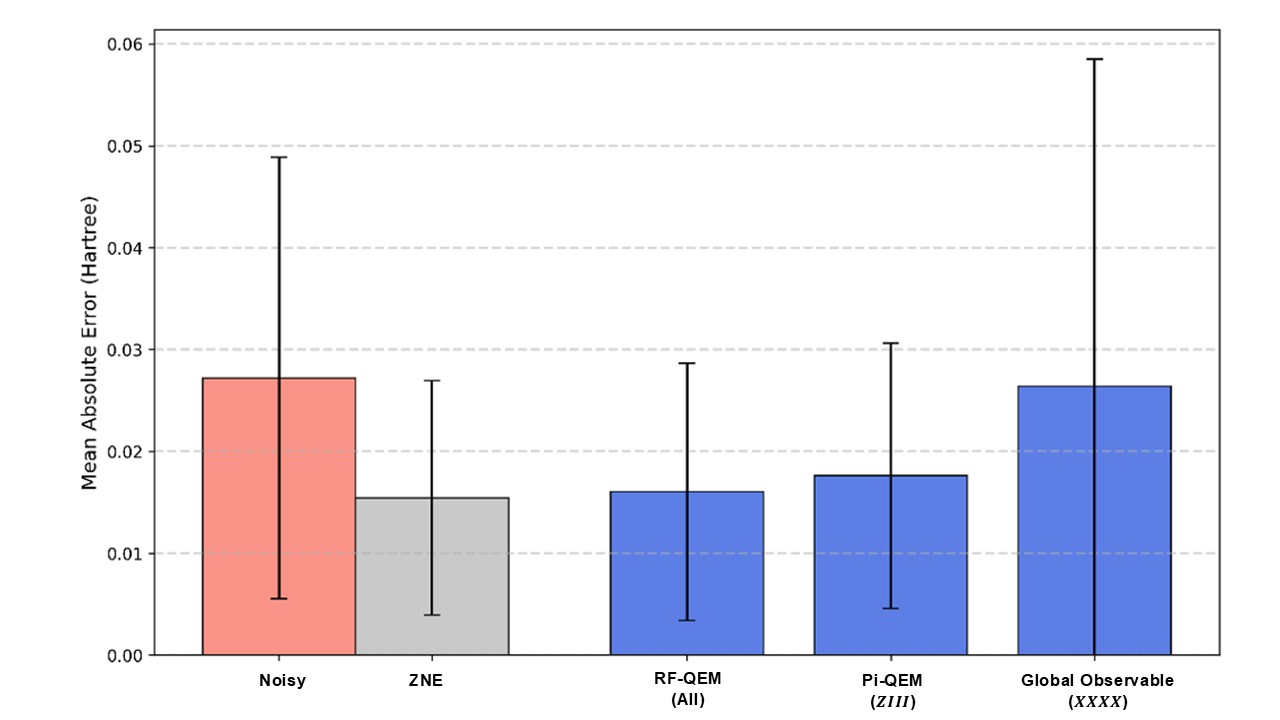}
    \caption{Mean absolute error (MAE) in ground-state energy estimation (Hartree). Unmitigated (noisy) and zero-noise extrapolation (ZNE) baselines are compared with standard RF-QEM, and Pi-QEM ($ZIII$). For contrast, a model trained on a global observable ($XXXX$) model is included. Error bars indicate the standard deviation across circuit configurations. Both RF-QEM and Pi-QEM achieve accuracies comparable to ZNE, whereas the global observable exhibits severe instability.}
    \label{fig:mae_energy_comparison}
\end{figure}

\begin{figure}[b]
    \centering
    \includegraphics[width=\columnwidth]{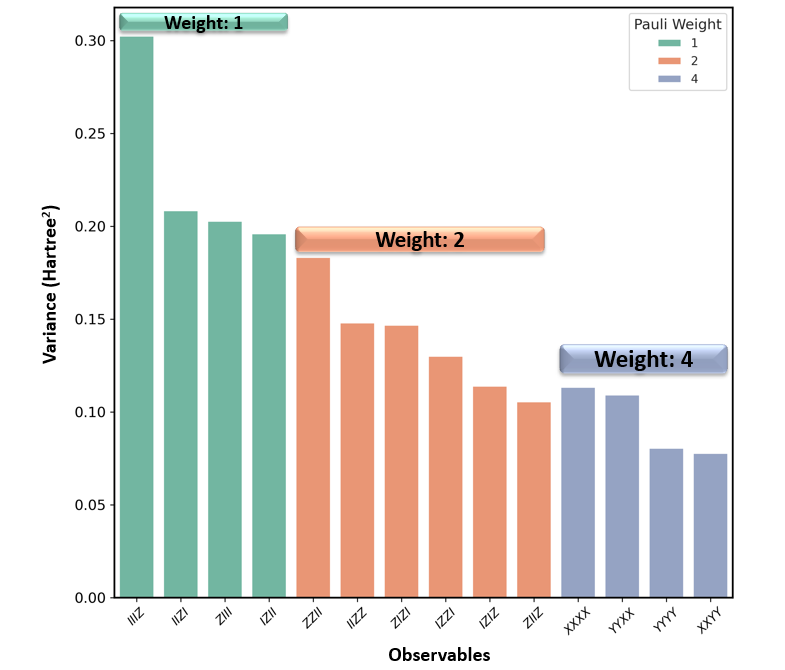}
    \caption{The variance distribution $\text{Var}_{\Theta} [f(P_k)]$ of individual Pauli strings over the PQC parameter space. The spectrum exhibits a clear hierarchy with a small number of observables dominating the variance and the majority contributing negligibly. Pi-QEM defines the selected subset $\mathcal{P}_{\text{sub}}$ using a Pauli-weight–based criterion that approximates the variance hierarchy of the observables. Observables with higher variance generate training datasets with greater statistical spread, enabling the regression trees ensemble to learn the noisy-to-ideal mapping more effectively. Furthermore, higher variance in the observables correlates with reduced MAE in Pi-QEM predictions, supporting the use of Pauli weight-based feature selection for efficient dataset construction. This also confirms that observables with higher variance and correspondingly low-weight local variables dominate the informative structure required for accurate mitigation. They are sufficient to capture the dominant noise structure governing the expectation values.}
    \label{fig:variance_distribution_hierarchy}
\end{figure}

Fig. 7 shows the MAE of the raw noisy measurement, digital ZNE, and models trained on the standard full set of Hamiltonians, a low-weight local observable $ZIII$ (Pi-QEM), and a high-weight global observable $XXXX$. The Pi-QEM model achieves an $\text{MAE} \approx 0.018$ Ha, which is statistically indistinguishable from the RF-QEM baseline ($\text{MAE} \approx 0.016$ Ha), despite using only a minimal subset. Furthermore, both standard RF-QEM and the Pi-QEM match the performance of ZNE ($\text{MAE} \approx 0.016$ Ha) while avoiding its associated circuit-folding overhead. In contrast, the model trained on the high-weight global observable ($XXXX$), which is associated with suppressed variance, yields an $\text{MAE} \approx 0.027$ Ha, representing a 22.90\% degradation in performance relative to the Pi-QEM model and approaching the unmitigated noisy baseline. 

Fig. 8 shows the variance distribution $\text{Var}_{\Theta} [f(\hat{P}_k )]$ for every Pauli string in the $\text{H}_2$ Hamiltonian. The spectrum admits a clear physical interpretation consistent with the cost-function-dependent barren plateau analysis \cite{Cerezo2021Cost}, where the variance is governed by the Pauli weight of each observable. Specifically, low-weight (local) observables occupy the top of the distribution, while high-weight (global) observables concentrate near the bottom due to variance suppression. In Fig. 8, the observable with the highest variance is $IIIZ$ ($\text{Var} \approx 0.31$, weight 1, where the weight corresponds to the number of non-identity Pauli operators). In contrast, the lowest-variance terms are the entangling operators $XXXX$, $YYYY$, $XXYY$, and $YYXX$ ($\text{Var} \approx 0.08$--$0.11$, all with weight 4). This ordering directly reflects the locality-dependent variance scaling predicted by barren plateau theory, where global observables exhibit significantly reduced variance. This structure provides a direct justification for Pi-QEM, as the variance hierarchy induces a natural ranking of observables by their statistical informativeness. Rather than explicitly thresholding the variance, Pi-QEM leverages its structural origin by selecting observables based on their Pauli weight, which reflects their locality and consequently their expected variance behavior. In this way, low-weight local observables are preferentially retained, while high-weight global observables associated with suppressed variance are systematically excluded. This criterion effectively filters out uninformative terms arising from concentrated landscapes, while preserving observables that encode meaningful parameter-dependent structure. In practice, the clear separation in the variance spectrum implies that a minimal subset, often a single, low-weight, local observable, is sufficient to achieve near-optimal performance.

\begin{figure}[b]
    \centering
    \includegraphics[width=\columnwidth]{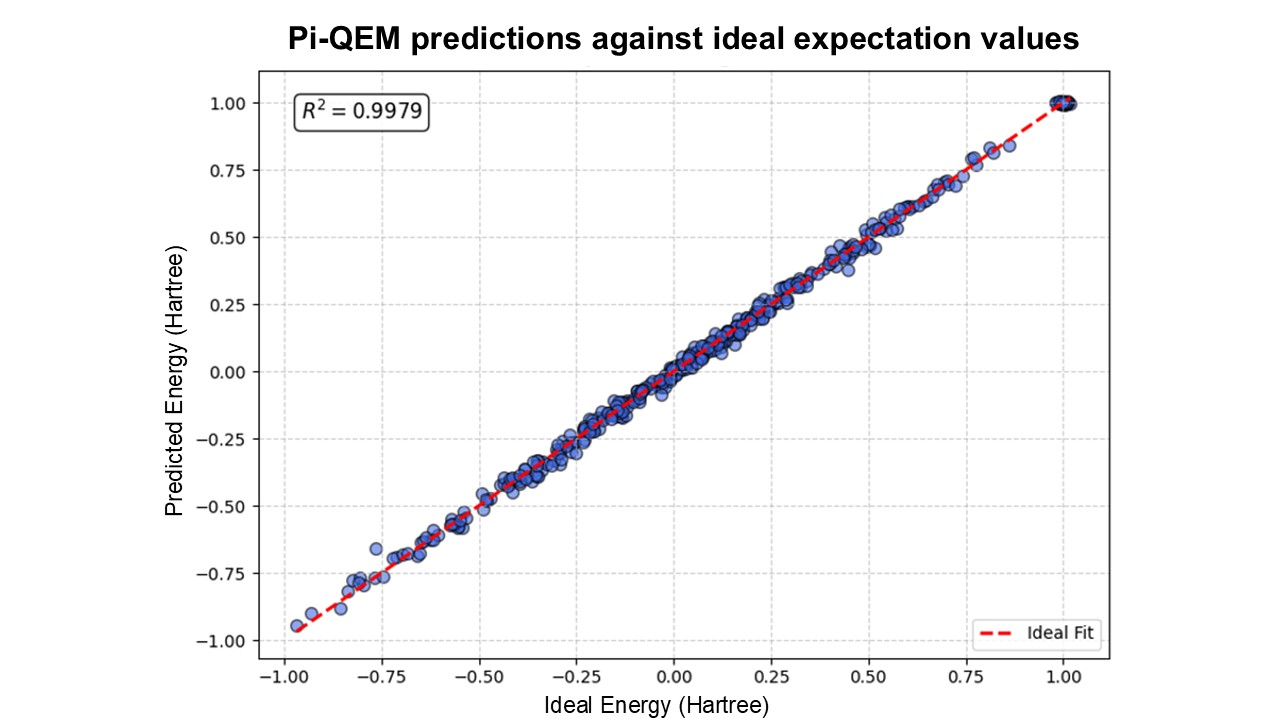}
    \caption{Scatter plots comparing Pi-QEM predictions with ideal expectation values. Data points cluster tightly along the ideal diagonal (red dashed line), yielding a high coefficient of determination ($R^2 \approx 0.9979$), demonstrating accurate learning of the nonlinear noise transformation.}
    \label{fig:scatter_predictions_ideal}
\end{figure}

\begin{figure*}[t]
    \centering
    \includegraphics[width=2.0\columnwidth]{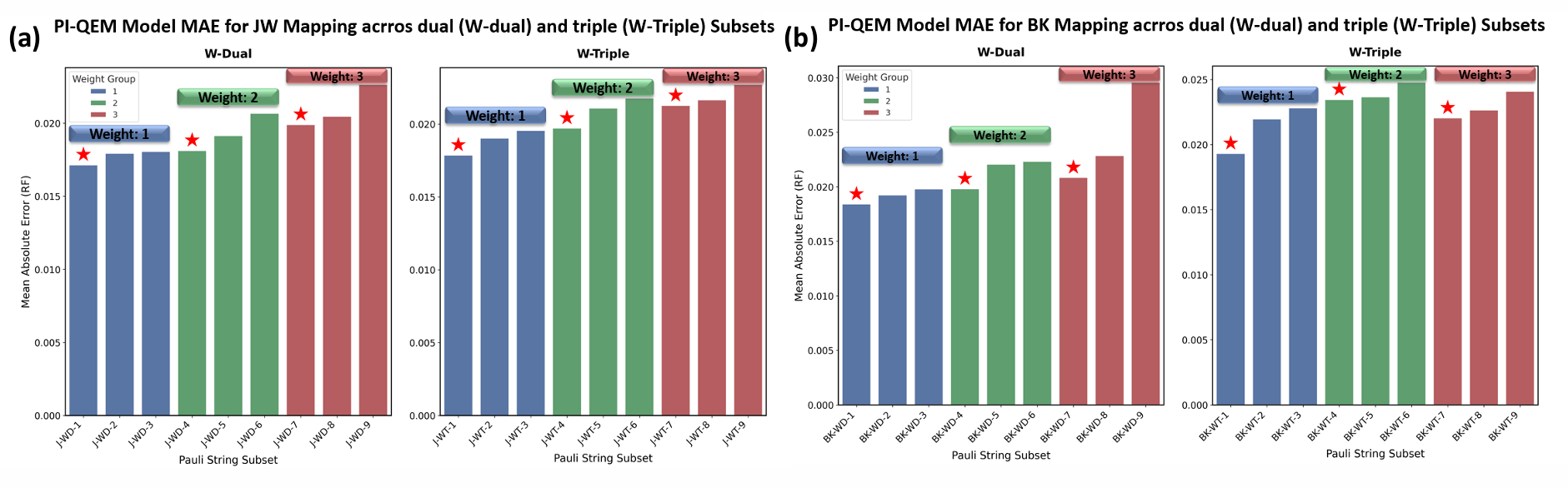}    
    \caption{Mean absolute error (MAE) of Pi-QEM models trained on different Pauli-string subsets for (a) Jordan–Wigner (JW) and (b) Bravyi–Kitaev (BK) mappings. W-Dual and W-Triple denote dual and triple observable subsets, respectively. Each bar represents a distinct subset configuration, and red stars indicate the optimal-performing subset within each group. Across all mappings, subsets composed of lower-weight (more local) observables consistently achieve lower MAE, while subsets including higher-weight (more global) terms exhibit degraded performance. These results demonstrate that, even for multi-observable configurations, mitigation performance is primarily governed by the locality of the selected observables.}
    \label{fig:mae_subsets_comparison}
\end{figure*}

\begin{table*}
\footnotesize
\setlength{\tabcolsep}{3pt} 
\caption{\label{tab:pauli_subsets_compact}
Pauli observable subsets used in Pi-QEM training, organized by feature cardinality (dual, triple) and Pauli-weight for the Jordan–Wigner (JW) and Bravyi–Kitaev (BK) mappings of the H$_2$ Hamiltonian. The subset codes (e.g., WD-1, WT-1) provided before the Pauli strings denote the unique identifiers for the Dual and Triple cardinality subsets, respectively.}
\begin{ruledtabular}
\begin{tabular}{c l l l l}
 & \multicolumn{2}{c}{Dual Cardinality} & \multicolumn{2}{c}{Triple Cardinality} \\
\cline{2-3} \cline{4-5}
Weight & Jordan–Wigner (JW) & Bravyi–Kitaev (BK) & Jordan–Wigner (JW) & Bravyi–Kitaev (BK) \\
\hline
1
 & WD-1: \textit{IIII}, \textit{ZIII} & WD-1: \textit{IIII}, \textit{IZII} & WT-1: \textit{IIIZ}, \textit{IIZI}, \textit{ZIII} & WT-1: \textit{IIIZ}, \textit{IIZI}, \textit{IZII} \\
 & WD-2: \textit{IIII}, \textit{IZII} & WD-2: \textit{IIII}, \textit{IIZI} & WT-2: \textit{IIIZ}, \textit{IZII}, \textit{ZIII} & - \\
 & WD-3: \textit{IIII}, \textit{IIZI} & WD-3: \textit{IIZI}, \textit{IZII} & WT-3: \textit{IIZI}, \textit{IZII}, \textit{ZIII} & - \\
\hline
2 
 & WD-4: \textit{IIZZ}, \textit{ZIZI} & WD-4: \textit{IIZZ}, \textit{ZIZI} & WT-4: \textit{IIZZ}, \textit{IZZI}, \textit{ZIZI} & WT-2: \textit{IIZZ}, \textit{IZIZ}, \textit{ZIZI} \\
 & WD-5: \textit{IIZZ}, \textit{ZZII} & WD-5: \textit{IZIZ}, \textit{ZIZI} & WT-5: \textit{IZZI}, \textit{ZIZI}, \textit{ZZII} & - \\
 & WD-6: \textit{ZIIZ}, \textit{ZZII} & WD-6: \textit{IIZZ}, \textit{IZIZ} & WT-6: \textit{IZIZ}, \textit{ZIIZ}, \textit{ZZII} & - \\
\hline
3
 & - & WD-7: \textit{IIZZ}, \textit{ZZIZ} & - & WT-3: \textit{IZIZ}, \textit{ZIZI}, \textit{IZZZ} \\
 & - & WD-8: \textit{IIZZ}, \textit{IZZZ} & - & WT-4: \textit{IIZZ}, \textit{ZZIZ}, \textit{IZZZ} \\
 & - & WD-9: \textit{ZIZI}, \textit{ZZIZ} & - & WT-5: \textit{IIZZ}, \textit{IZIZ}, \textit{ZZIZ} \\
\hline
4
 & WD-7: \textit{YYYY}, \textit{XXXX} & WD-10: \textit{ZZZI}, \textit{ZZZZ} & WT-7: \textit{XXYY}, \textit{YYXX}, \textit{XXXX} & WT-6: \textit{ZXIX}, \textit{ZXZX}, \textit{IXIX} \\
 & WD-8: \textit{XXYY}, \textit{YYXX} & WD-11: \textit{ZXIX}, \textit{ZXZX} & WT-8: \textit{YYYY}, \textit{XXYY}, \textit{XXXX} & WT-7: \textit{ZXIX}, \textit{IXZX}, \textit{ZXZX} \\
 & WD-9: \textit{YYXX}, \textit{XXXX} & WD-12: \textit{IXZX}, \textit{ZXZX} & WT-9: \textit{YYYY}, \textit{XXYY}, \textit{YYXX} & WT-8: \textit{IXZX}, \textit{ZXZX}, \textit{IXIX} \\
\end{tabular}
\end{ruledtabular}
\end{table*}

Fig. 9 compares the predicted expectation values produced by the Pi-QEM model against the ground-truth ideal values. The predictions align closely with the diagonal, indicating excellent agreement with the ideal results. Quantitatively, the coefficient of determination satisfies $R^2 \approx 0.998$, demonstrating that nearly all variance in the ideal expectation values is captured by the trained Pi-QEM model.

Fig. 10 investigates whether the Pi-QEM selection principle generalizes across different fermion-to-qubit mappings, namely the Jordan-Wigner (JW) and Bravyi-Kitaev (BK) transformations, and whether increasing the feature cardinality from single to dual and triple configurations further improves mitigation performance. Therefore, additional dataset categories are considered: dual and triple, corresponding to the number of Pauli strings used as features. This approach allows a detailed analysis of the contribution of individual Pauli strings to the mitigation performance and facilitates the identification of the most informative observables. The specific Pauli observable subsets employed for each mapping and feature cardinality are summarized in Table 1, which lists the selected observables and their corresponding dual and triple combinations according to their Pauli weights.

The MAE shown in Fig. 10 reveals a clear structural pattern consistent with the cost-concentration theory of PQCs, and extends this observation to multi-observable subsets where a clear weight-dependent trend emerges for both W-Dual and W-Triple configurations. For W-Dual, subsets composed of lower-weight observables consistently achieve the best performance, with weight-1 combinations (e.g., JW-WD-4, JW-WD-3, JW-WD-2) yielding an $\text{MAE} \approx 0.017$--$0.018$. As the Pauli weight increases, the error rises systematically: weight-1 subsets reach $\approx 0.018$--$0.019$, weight-2 subsets $\approx 0.019$--$0.020$, and weight-3 subsets show a marked degradation to $\approx 0.025$--$0.026$. The same trend is observed for W-Triple configurations, where weight-1 subsets achieve an $\text{MAE} \approx 0.0179$--$0.0197$, followed by weight-2 subsets at $\approx 0.0195$--$0.0205$, and a significant performance drop for weight-3 subsets, reaching up to $\approx 0.030$. This weight-dependent hierarchy is further validated by the BK mapping, which includes an additional weight-3 group. In the BK W-Dual setup, MAE scales predictably from $\approx 0.018$–$0.020$ for weight-1, $\approx 0.020$–$0.022$ for weight-2, $\approx 0.022$–$0.023$ for weight-3, and up to $\approx 0.027$ for weight-4 (with BK-WD-10 acting as a low-error outlier within its group). The BK W-Triple configuration mirrors this progression, scaling from $\approx 0.019$ (weight-1) up to $\approx 0.026$ (weight-4).

These results align with the subset constructions in Table 1 and are consistently observed across all mappings, indicating that the behavior is independent of the encoding scheme. Overall, the results show that mitigation performance in multi-observable settings is strongly governed by Pauli weight. Subsets composed of local (low-weight) observables consistently outperform those containing global (higher-weight) terms, providing strong empirical support for the Pi-QEM principle that statistical informativeness is concentrated in low-weight observables.

\section{Conclusion}

In this work, we introduce Pauli weight Quantum Error Mitigation (Pi-QEM), a systematic observable selection framework for learning-based quantum error mitigation. By using the variance $\text{Var}_{\Theta} [f(\hat{P}_k)]$ as a measure for statistical informativeness, Pi-QEM provides a theoretically grounded criterion for constructing efficient training datasets for regression tree models, while implementing this criterion in practice through a Pauli-weight–based selection rule.

We established the connection between observable variance and the cost-concentration phenomenon in PQCs, showing that variance directly reflects the locality-dependent barren plateau structure. High-variance observables correspond to regimes with nontrivial parameter dependence and preserved training signal, while low-variance observables are associated with concentrated landscapes that yield uninformative data. This insight allows Pi-QEM to systematically select informative observables based on their Pauli weight, retaining low-weight (local) terms while excluding high-weight (global) ones associated with suppressed variance.

Our numerical simulations of the $\text{H}_2$ molecule under the JW mapping demonstrate that the variance spectrum exhibits a clear hierarchy governed by Pauli weight, consistent with theoretical predictions. Exploiting this structure, we show that a minimal Pi-QEM-selected subset, often a single dominant low-weight local observable, is sufficient to achieve mitigation performance comparable to the full-Hamiltonian standard RF-QEM baseline. In particular, we observe up to a 34.01\% reduction in ground-state energy error, while reducing the training set size from $\mathcal{O}(MN)$ to $\mathcal{O}(M|\mathcal{P}_{\text{sub}}|)$, with $|\mathcal{P}_{\text{sub}}| \ll N$. 

Furthermore, training on high-weight global observables associated with suppressed variance leads to significant degradation, confirming that observable variance, captured in practice through Pauli weight, is a key requirement for effective learning-based error mitigation. More broadly, these findings highlight the importance of aligning machine learning strategies with the underlying geometry of quantum optimization landscapes, opening a pathway toward resource-efficient error mitigation in near-term quantum algorithms.

\section{Code Availability}
The source code implementing the PiQEM framework is publicly available at:
\url{https://github.com/fadhil698/PiQEM}

\bibliography{references}
\end{document}